\documentclass[manuscript]{aastex}
\usepackage{epsfig}
\usepackage{natbib}
\usepackage{subfigure}
\usepackage{bigints}
\usepackage{amssymb,amsmath}

\DeclareMathVersion{bold}

\newcommand{\eg}{e.\,g.}
\newcommand{\dmunits}{{\rm pc\,cm^{-3}}}

\newcommand{\sact}{S_{\rm actual}}
\newcommand{\tsky}{T_{\rm sky}}

\newcommand{\tsys}{T_{\rm sys}}

\newcommand{\thorn}{T13}
\newcommand{\ti}{w}
\newcommand{\wfrb}{w_{\rm frb}}
\newcommand{\tI}{t_{\rm eff}}
\newcommand{\taus}{\tau}

\newcommand{\tsamp}{t_{\rm s}}

\newcommand{\vf}{\mbox{V-FASTR}}

\newcommand{\startdate}{24 Apr 2011}
\newcommand{\finaldate}{24 Feb 2015}

\shorttitle{Limits on Fast Radio Bursts from Four Years of V-FASTR}
\shortauthors{Burke-Spolaor et al.}

\begin{document} 

\title{Limits on Fast Radio Bursts from Four Years of the V-FASTR Experiment}

\author{S. Burke-Spolaor$^{1,2,3}$}
\author{Cathryn M. Trott$^{4,7}$}
\author{Walter F. Brisken$^{1,5}$}
\author{Adam T. Deller$^{6}$}
\author{Walid A. Majid$^{3}$}
\author{Divya Palaniswamy$^{4}$}
\author{David R. Thompson$^{3}$}
\author{Steven J. Tingay$^{4,7}$}
\author{Kiri L. Wagstaff$^{3}$}
\author{Randall B. Wayth$^{4,7}$}
\affil{$^1$NRAO, P.O. Box O, Socorro, NM 87801, USA; sspolaor@nrao.edu}
\affil{$^2$California Institute of Technology, 1200 E California Blvd, Pasadena, CA 91125, USA}
\affil{$^3$Jet Propulsion Laboratory, California Institute of Technology, Pasadena, CA 91109, USA}
\affil{$^4$ICRAR, Curtin University, Bentley, WA 6845, Australia}
\affil{$^5$University of Minnesota, MN 55155, USA}
\affil{$^6$ASTRON, Oude Hoogeveensedijk 4, 7991 PD Dwingeloo, The Netherlands}
\affil{$^7$ARC Centre of Excellence for All-Sky Astrophysics (CAASTRO)}

\begin{abstract}
The \vf\ experiment on the Very Long Baseline Array was designed to detect dispersed pulses of milliseconds duration, such as fast radio bursts (FRBs). We use all \vf\ data through February 2015 to report \vf's upper limits on the rates of FRBs, and compare these with re-derived rates from Parkes FRB detection experiments. \vf's operation at $\lambda=20$\,cm allows direct comparison with the 20\,cm Parkes rate, and we derive a power-law limit of $\gamma<-0.4$ (95\% confidence limit) on the index of FRB source counts, $N(>$$S)\propto S^\gamma$. Using the previously measured FRB rate and the unprecedented amount of survey time spent searching for FRBs at a large range of wavelengths ($0.3\,{\rm cm}>\lambda>90$\,cm), we also place frequency-dependent limits on the spectral distribution of FRBs. The most constraining frequencies place two-point spectral index limits of $\alpha_{\rm 20\,cm}^{\rm 4\,cm}<5.8$ and $\alpha_{\rm 90\,cm}^{\rm 20\,cm}>-7.6$, where fluence $F\propto f^\alpha$ if we assume true the burst rate reported by \citet{championbursts} of $R(F\sim 0.6\,{\rm Jy\,ms}) = 7\times10^3\,{\rm sky^{-1}\,day^{-1}}$ (for bursts of $\sim$3\,ms duration). This upper limit on $\alpha$ suggests that \emph{if} FRBs are extragalactic but non-cosmological, that on average they are not experiencing excessive free-free absorption due to a medium with high optical depth (assuming temperature $\sim$8,000\,K), which excessively invert their low-frequency spectrum. This in turn implies that the dispersion of FRBs arises in either or both of the intergalactic medium or the host galaxy, rather than from the source itself.
\end{abstract}

\section{Introduction}\label{sec:intro}
In the past decade, we have gathered the first perspicuous evidence that there exists a large population of ``fast radio bursts'' (FRBs) with a potentially extragalactic origin \citep[\eg][]{LB,thornton13}.
However, uncertainties remain as to the origins of FRBs, and despite the large implied sky rate of $\sim$$10^4/{\rm sky/day}$, there have been only $\sim$20 detections reported to date \citep{thornton13,spitlerFRB,bannisterbursts,ravishannonFRB,petroffrealtime,championbursts}, largely from Parkes Radio Telescope. Other experiments have experienced non-detections which, so far, are consistent with the rate and sensitivity of detections \citep{flyseye,lofarlimits,caseyVLA}. Mounting evidence points to an extragalactic origin; namely:
1) the dispersion measures and rotation measures in excess of those expected from the Milky Way, and scattering timescales much smaller than those expected from a source of Galactic origin, imply that propagation through extragalactic media (from the intergalactic medium, a host galaxy, or the source itself) imparts these values \citep{luan,kulkarni14,masui};
and 2) the absence of FRBs detections at low Galactic latitudes implies that ``filtering'' and signal dampening effects may be arising from propagation through the Milky Way \citep{bannisterbursts,petroffgb}. Recent detections and debate have added further complexity to the story: one FRB has been detected multiple times \citep{repeatingFRBnature}. Another publication reported the first potential afterglow candidate to an FRB, implying that at least that FRB might be arriving from a galaxy at redshift $z=0.49$ \citep{keanenatureFRB}, although whether the afterglow is truly associated with the FRB event is still under debate \citep{wbarxiv,wbatel,vedantham16}.


Since 2011, the \vf\ experiment has run commensally on the Very Long Baseline Array (VLBA), which can provide a few tens of milli-arcsecond localization of any detected FRB---hence the unique potential to localize an extragalactic pulse to a precise location within a host. The experiment has been described in detail in two publications \citep{wayth11,thompson11}, and the sensitivity of the experiment has been reported twice at earlier stages of observation \citep{wayth12,trott13}. In brief, \vf\ uses an incoherent sum of the correlated data from the VLBA antennae to dedisperse and search for bursts in the resulting time series. For any promising candidate above our search threshold, we are able to access the baseband data and image the candidate to confirm it as an FRB. \vf\ is one of few FRB detection experiments currently operating worldwide at a broad range of frequencies, and thus has the potential to provide unique limits on average FRB population features such as spectral index and number count scaling. This is a particularly strong capability given the rapid and huge spectral index range reported for the repeating source of \citep{repeatingFRBnature}.

Since its campaign commenced, \vf\ has not yet detected any FRBs.
This paper provides a careful analysis of the sensitivity limits of the \vf\ data taken to date, and interprets the non-detection in terms of the physical parameters of FRBs with respect to previous FRB detection rates.

\begin{table*}
\centering
  \caption{Observing parameters for the \vf\ data as of \finaldate.
}\label{table:nhours}
\begin{tabular}{lcccccr}
\hline
{\bf Band}& {\bf $\langle f_{\rm ctr}\rangle$}\tablenotemark{a} & {\bf SEFD\tablenotemark{b}} & {\bf SEFD$_{\rm eff}$} & {\bf $\tsys$}\tablenotemark{c} & &  {\bf $N_{\rm hrs}$}\\
{\bf (cm)} & (GHz) & {\bf(Jy)} & {\bf(Jy)} & {\bf (K)} & $\langle N_{\rm ant}\rangle^{\rm a}$ & {\bf (h)}\\
\hline
90 & 0.318 & 2742 & 3439 & 184 & 9.2 & 24.2 \\
50 & 0.465 & 2744 & 3126 & 206 & 9.3 & 10.2 \\
20 & 1.550 & 302\tablenotemark{d} & 311 & 31\tablenotemark{d} & 9.4 & 1648.0 \\
13 & 2.278 & 347 & 357 & 30 & 9.0 & 82.8 \\
13/4\tablenotemark{e} 
   & 5.537 & 399 & 400 & 37 & 9.0 & 491.8 \\
6 & 5.949 & 244\tablenotemark{d} & 245 & 28\tablenotemark{d} & 9.1 & 1264.9 \\
4 & 8.418 & 327 & 327 & 36 & 10.1\tablenotemark{f} & 1426.8 \\
2 & 15.082 & 543 & 543 & 67 & 9.6 & 790.0 \\
1 & 22.312 & 640\tablenotemark{g} & 640 & 68\tablenotemark{g}& 9.6 & 1493.8 \\
0.7 & 43.161 & 1181 & 1181 & 106 & 9.3 & 797.7 \\
0.3 & 86.312 & 4236 & 4236 & 119 & 7.8 & 115.9 \\
\hline
\hline
\tablenotetext{a}{Observing-time-weighted average.}
\tablenotetext{b}{\mbox{\scriptsize https://science.nrao.edu/facilities/vlba/docs/manuals/oss/bands-perf} accessed 2015-04-20. We assume a frequency-independent value for each band.}
\tablenotetext{c}{Averaged across antennae, from internal system measurements file on 2015-04-20. We assume a $f$-independent value for each band.}
\tablenotetext{d}{Average of the two values reported for this receiver.}
\tablenotetext{e}{\vf\ treats the dichroic ``S/X band'' receiver observations as a single band when correcting for dispersion; here we quote the average values between this receiver's S and X bands.}
\tablenotetext{f}{Some observations include non-VLBA antennae, \eg\ the GBT, tied-array VLA, or EVN antennae. This occurs most frequently in the 4\,cm band for geodesy experiments.}
\tablenotetext{g}{We use the values quoted for 1\,cm, which is closest to our $\langle f_{\rm ctr}\rangle$.\vspace{0mm}}
\end{tabular}
\end{table*}

\section{Four Years of V-FASTR}\label{sec:system}
\subsection{Data: \startdate\ to \finaldate}
As a commensal experiment, the observing set-up of \vf\ data is distinctively inhomogeneous. This is reflected by the values populating our 5-dimensional observing parameter distribution, represented in Table \ref{table:nhours} and Figure \ref{fig:data}. The latter demonstrates that \vf\ searches for FRBs at the full range of VLBA operating frequencies, and at a broad range of total bandwidth and channel width, both of which depend on the experiment on which \vf\ is piggybacking. The range of time sample lengths reflects the fact that \vf\ splits off a parallel data stream, for which the sampling time is optimized depending on the dispersion smearing at the given frequency set-up of the primary experiment.
Table \ref{table:nhours} columns are as follows: (1) the observing receiver, as given by a representative wavelength; (2) the average center frequency of the \vf\ data for that band; (3) the system equivalent flux density (SEFD) quoted for the VLBA; (4) the effective SEFD as described in \S\ref{sec:sefd}; (5) the system temperature; (6) the average number of antennae used for the \vf\ search; and (7) the cumulative number of hours spent using that receiver. 
Below we describe the dedicated V-FASTR observing program BT127, as well as the data removal heuristics we used to remove observing scans (from all observations) that are unsuitable for use in our limit analysis.

\subsection{BT127 targeted observing program}
The one exception to \vf's commensal operation has been a recent targeted program (BT127, which commenced in 2014) approved under the ``VLBA Filler Project Challenge''\footnote{https://science.nrao.edu/enews/7.1/}, in which projects can obtain large amounts of VLBA observation time in the gaps between long-duration observations.  Such a filler mode is ideal for \vf, as a list of high ($|b|\gtrsim20$) Galactic latitude targets can be defined at any right ascension (RA).

We constructed a list of 48 target fields spaced in RA by 0.5\,h.  All fields lie north of the equator.  Twenty nine of these fields are at high Galactic latitudes ($|b|>20^\circ$) and utilise observations that switch between the selected field and a high quality calibrator on a ten minute cadence.  For the remaining 19 right ascensions for which a high Galactic latitude selection with nearby high-quality calibrator was not possible, we reverted to performing continuous observations of a field containing a strong calibrator near a declination of $+30^{\circ}$.

BT127 observations are scheduled for the object most appropriate for the time range covered by an available filler slot.  BT127 observations utilize the full bandwidth available at 1.4\,GHz in order to achieve maximum sensitivity.  As of \finaldate, a total of $\sim$350\,h of observation have been obtained via BT127. 
However, not all of these data were processed by \vf, and additional filtering of the data is described below. Therefore, 158 hours of BT127 data have been usable for the analysis presented in this publication.


\subsection{Invalid data}\label{sec:baddata}
From all data used by \vf\ (including all projects, not just BT127), some were considered unusable due to the presence of a large number of spurious candidates---hence the manual inspection step was not feasible or careful for these observations---or because the observing system had severe errors and did not record usable data. We thus do not consider data as follows:
\begin{itemize}
\item Observations centered on either of the bright pulsars PSR\,J0332+5434 or PSR\,J1136+1551.
\item Projects processed between 30 Jan and 4 Feb 2014 due to a communications error between the correlator and \vf\ processing disk.
\item Projects processed between 15 and 24 July 2014 due to a temporary software error.
\end{itemize}
The invalid observations, in total encompassing 148\,h (1.8\%) of the \vf\ data, are not considered in our analysis.

\begin{figure*}
\centering
\includegraphics[width=0.485\textwidth,trim=19mm 21mm 7mm 21mm, clip]{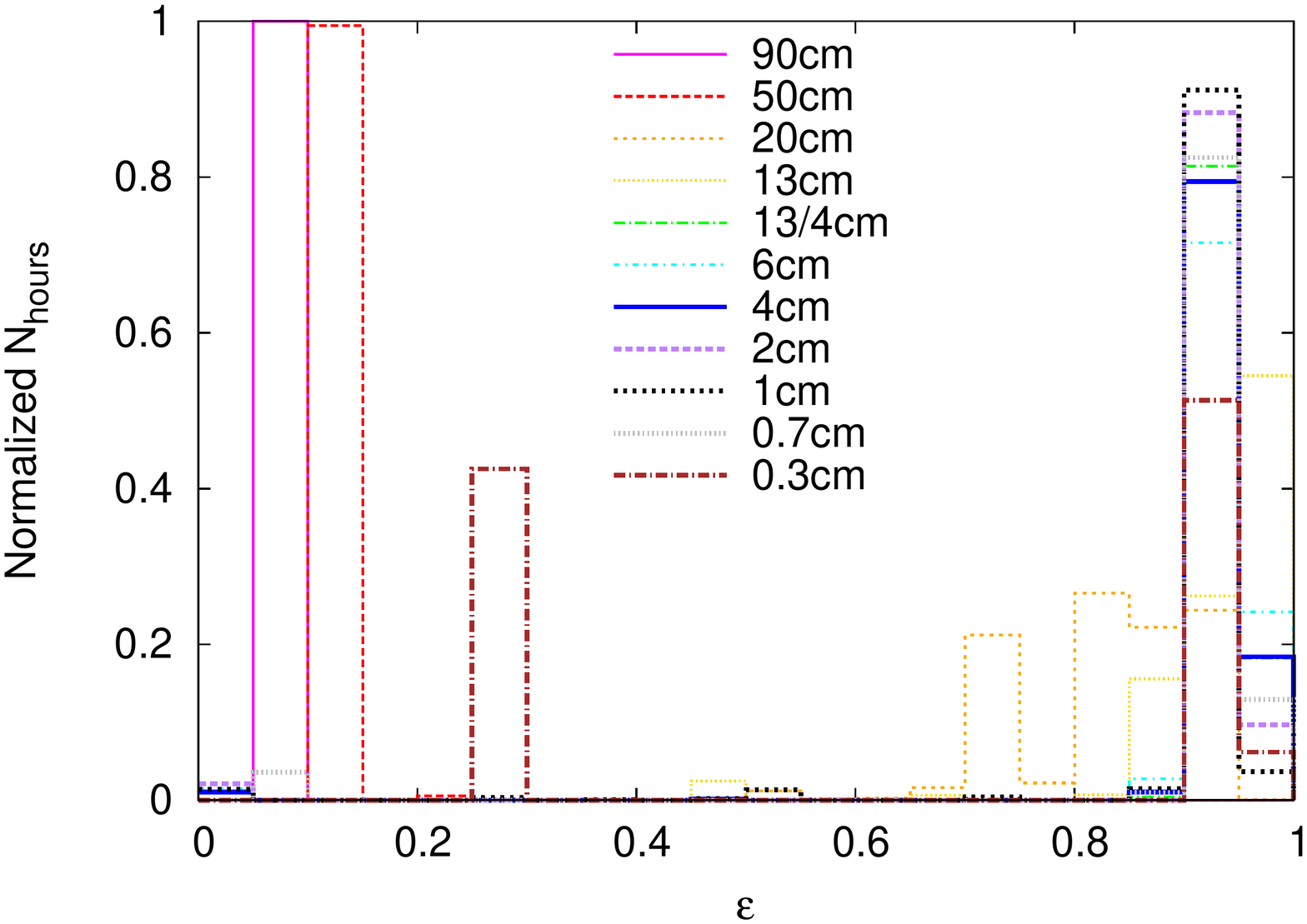}
\includegraphics[width=0.485\textwidth,trim=19mm 21mm 7mm 21mm, clip]{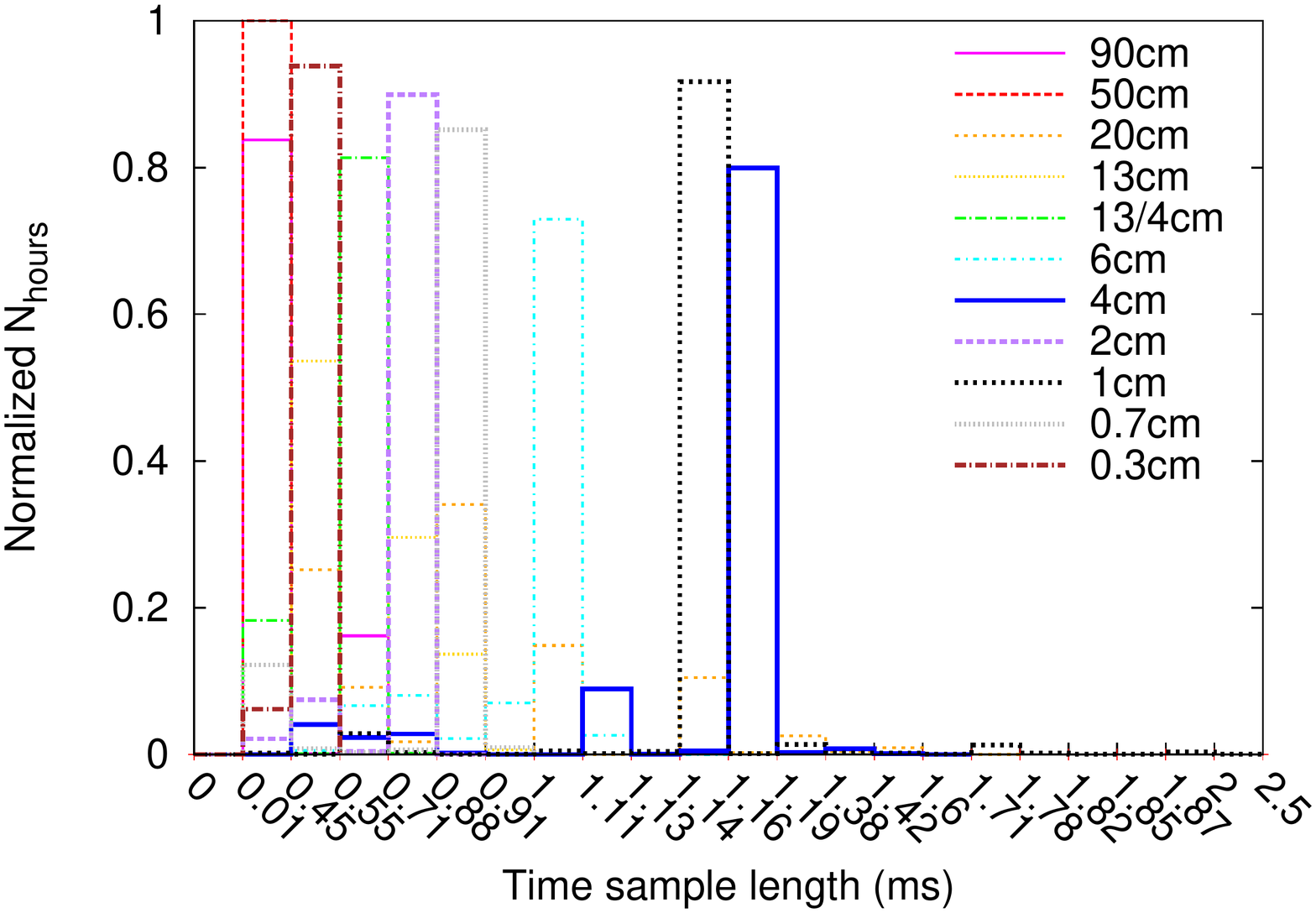}
\includegraphics[width=0.485\textwidth,trim=19mm 21mm 7mm 21mm, clip]{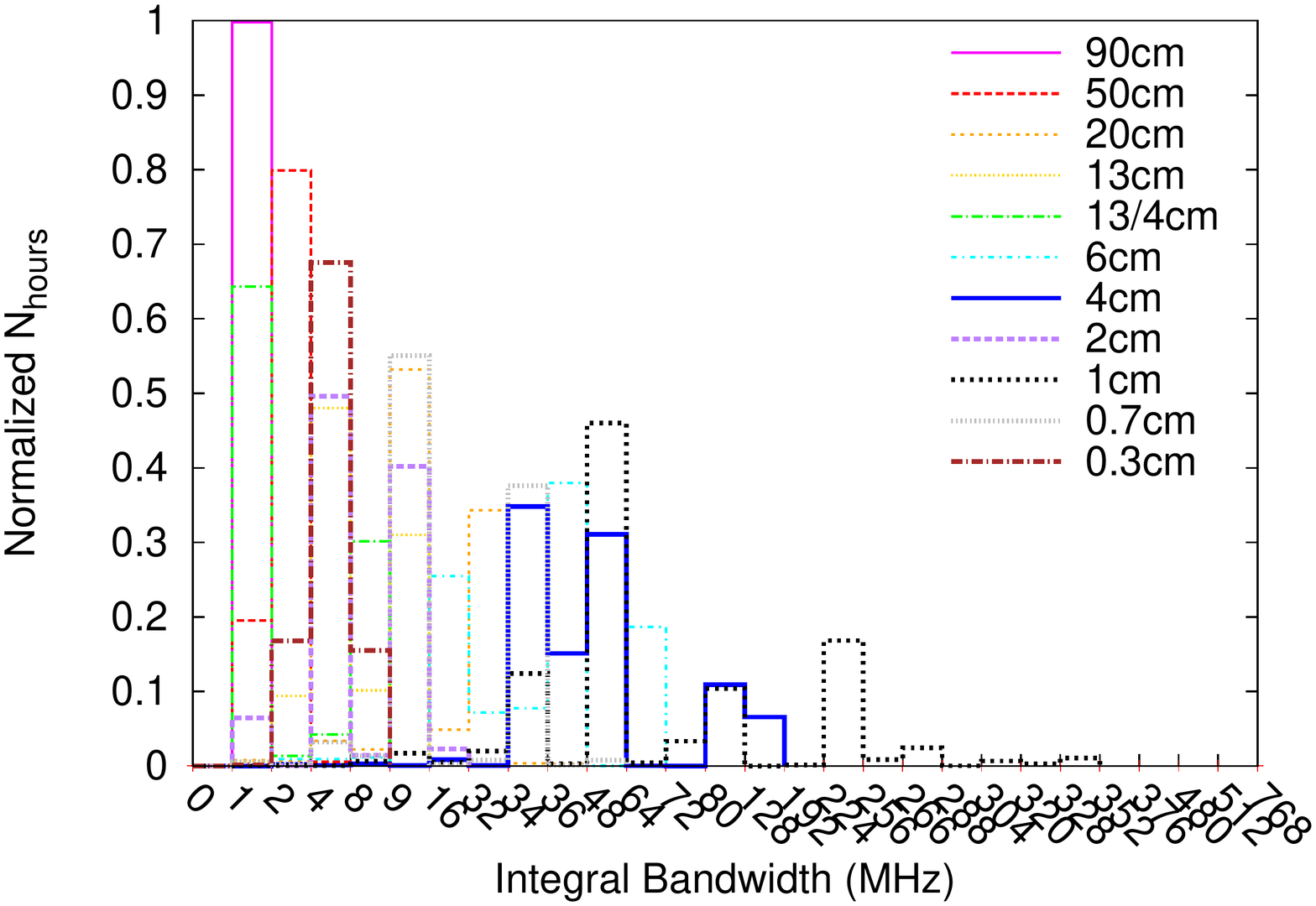}
\includegraphics[width=0.485\textwidth,trim=19mm 21mm 7mm 21mm, clip]{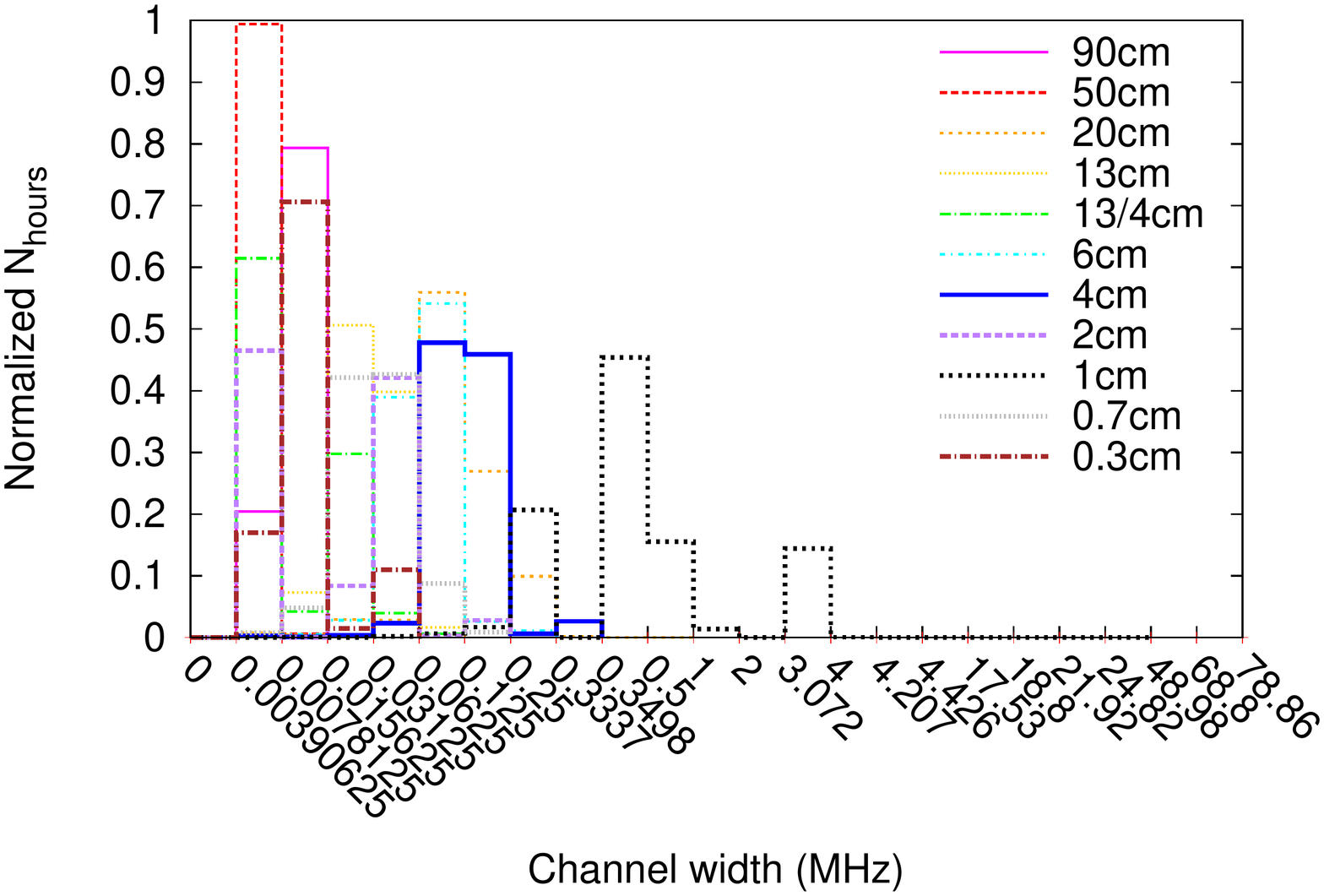}
\caption{Two-dimensional projection (Fraction of time in a given band spent at different degradation factors, sampling times, channel widths, and bandwidths) of the five-dimensional distribution contributing to our sensitivity curve calculations (\S\ref{sec:limitplots}), demonstrative of the inhomogeneity in the \vf\ data distribution. 
}\label{fig:data}
\end{figure*}

\section{Limit methodology and V-FASTR Sensitivity Considerations}\label{sec:prelim}
The limits derived in this paper are based on the framework of \citet[][]{trott13}, which takes into account the probabilistic nature of detection and transient event occurrence, enabling the construction of 2D probability distributions in the sensitivity - rate parameter space. That work provides a full description of the framework's derivation, including an initial application of the framework specifically to the \vf\ experiment. We provide here the relevant equations and alterations of their framework used for our analysis. Where elements are not explicitly expanded (\eg\ $B(f,\theta)$ below), we refer the reader to \citealt{trott13}.
The Trott et al.\ ``detection performance metric,'' which casts sensitivity factors to a common reference frequency $f_0$, is given by:
\begin{equation}
\epsilon S_{\rm actual}(f_0,\theta)\sqrt{\ti} =
\frac{C}{B(\theta)\Big (\frac{f}{f_0}\Big)^\alpha}\sqrt{\frac{S_{\rm sys}^2(f)~\tsamp}
{N_{\rm ant}}}~.
\label{eq:sens}
\end{equation}
Here, $\sact$ and $\ti$ are the actual source flux density and intrinsic pulse width, unobstructed by scattering and any observing instrumentation. Our detection degradation factor $\epsilon$ differs from that given by Trott et al., and is described in Sec.\,\ref{sec:broadening}. The factors $\tsamp, C,$ and $N_{\rm ant}$ give the sampling time, the signal-to-noise detection threshold (a dimensionless multiplier of the noise), and the number of antennas summed for detection, respectively. $S_{\rm sys}$ is the frequency-dependent system sensitivity; while it varies for different receivers, we use a fixed value for each frequency band (SEFD$_{\rm eff}$ as described in \S\ref{sec:sefd}). $B(\theta)$ is the antenna pattern. The transient signal itself might be frequency dependent, represented here by the power-law spectral index $\alpha$. 

A source with true flux density $S_{\rm actual}$ is measured with a flux density that includes statistical noise, which is Gaussian-distributed with mean $\mu=S_{\rm actual}$ and data variance $\sigma^2$. The probability a source of flux density $S_{\rm actual}$ will be detected above a threshold given by $C\sigma$ is given by the cumulative distribution function:
\begin{equation}\label{eq:plim}
P(S_{\rm actual}>C\sigma) = \bigintssss_{C\sigma}^\infty\mathcal{N}(S_{\rm actual},\sigma^2)dS = \frac{1}{2}+\frac{1}{2}{\rm erf}\bigg ( \frac{S_{\rm actual}-C\sigma}{\sqrt(2)\sigma} \bigg )~,
\end{equation}
where $\mathcal{N}(\mu,\sigma^2$) denotes a Gaussian distribution. The cumulative probability in the rightmost part of the equation describes the exclusion regions. The cumulative probability that an event is not detected (due to noise) is the complementary function, $1-P(S_{\rm actual}> C\sigma)$ \citep{trott13}.

The 2D probability distributions are formulated as a test of the alternate hypothesis, $1-P_{\rm null}$, where the null hypothesis is that at least one source would have been detected when $n$ events have occurred:
\begin{equation}\label{eq:pnull}
P_{\rm null} = \sum_{n>1}\Big([1-P(S_{\rm actual}<C\sigma)^n]~\times~Q(n;\lambda)\Big)~.
\end{equation}
Here, $Q(n;\lambda)$ is a Poisson distribution, \mbox{$\lambda/({\rm FOV}\times \mathcal{T}_{\rm obs})$} is the expected number density of events per field of view and total observing time, and $C\sigma$ refers to a detection threshold of $C$ times the noise, $\sigma$. The alternate hypothesis is the probability that events that occurred were not detectable by \vf.
The net probability for $N$ independent experiments is
\begin{equation}\label{eq:ptot}
P_{\rm tot} = \prod^N_{i=1} [1-P_{\rm null}, i]
\end{equation}
In this way, any number of inhomogeneous experiments can be combined into a single constraint on the rates of millisecond transients in the phase space of ($\epsilon S_{\rm actual} \sqrt{\ti}$, FOV$\times T_{\rm obs}$), which represent the true flux and the true rate of bursts.

To summarize this process for clarity: in order to compute Eq.~\ref{eq:ptot} for a given instrumental set-up, we use the  observational parameters represented in Fig.\,\ref{fig:data} to calculate the sensitivity thresholds for Eq.~\ref{eq:plim} as given by Eq.~\ref{eq:sens}. Each instrumental set-up has a given field of view and on-sky time, and these are used to calculate $Q(n;\lambda)$ in Eq.~\ref{eq:pnull} above. We then use Eq.~\ref{eq:ptot} to combine the probabilities from various instrumental set-ups into the probability that \vf\ would have detected a burst of a given true flux $S_{\rm actual}\sqrt{w}$ and true rate FOV$\times T_{\rm obs}$.

\subsection{SEFD and Sky Temperature}\label{sec:sefd}
In our analysis, we take a band-averaged system sensitivity $S_{\rm sys}$ to be the effective system equivalent flux density (SEFD$_{\rm eff}$) for the VLBA. 
The quoted SEFD of the VLBA is reported in Table \ref{table:nhours}, and we estimate SEFD$_{\rm eff}$ as reported below. The SEFD and $\tsys$ values reported in Table \ref{table:nhours} correspond to a fairly optimistic observing configuration, and may be further affected by telescope elevation, atmospheric sky temperature, and interference. Sensitivity values in typical observations can be on average $<$20\% worse at the lowest and highest frequencies, and roughly equal to those reported in this table for intermediate GHz frequencies (the most stable $\tsys$ values). The effects of harmful interference on the SEFD are mitigated by its automatic removal performed in the \vf\ pipeline. 

Notable amounts of time are spent observing the Galactic plane, at which---particularly at lower frequency---the background sky temperature can be comparable to the receiver temperature, and therefore represent a significant loss in sensitivity. Thus the effective SEFD for a pointing is a sky-position and frequency-dependent scaling of the SEFD:
\begin{equation}
{\rm SEFD_{\rm eff}}(f,l,b) = {\rm SEFD}\cdot\frac{\tsky(f,l,b)+\tsys}{\tsys}~.
\end{equation}
We determined $\tsky$ from the observation-based model of \cite{globalskymodel}. A representative $\tsky$ for an observation is taken as the observing-time-weighted average of $\tsky(l,b)$ at the highest and lowest  observed frequencies. 
It is clear that for \vf\ observations to date, background sky temperature is only a notable factor at the 2\,GHz band and lower.

\subsection{The \vf\ Detection Threshold}
\label{sec:rfi-C}
The signal-to-noise detection threshold value, $C$, is nominally defined as a multiplication over the thermal noise in the receiver, however due to the presence of radio frequency interference (RFI), $C$ can represent a larger sensitivity value. For millisecond pulse searches, typical values range $C=6$ on the low end and $C=10$ on the high end \citep[\eg][]{RRATs,thornton13,spitlerFRB,bannisterbursts}. The impact of RFI on $C$ has, to our knowledge, not yet been considered in any FRB sensitivity analysis for single-dish experiments. \vf\ has some intrinsic RFI advantage over single dish experiments, as the spatially disparate antennae form a natural anti-coincidence filter \cite[\eg][]{thompson11}. The \vf\ processing pipeline also performs adaptive RFI excision of narrow- and broad-band signals as described in \citet{wayth12}, which removes the bulk of potent RFI. Severely sensitivity-impacting narrow- and broad-band signals have been seen at the manual inspection stage in only a negligible fraction of the newer broad-band data. 

V-FASTR is unique in that it incorporates multiple stations, each of which suffers different location-dependent RFI.  It accounts for this dynamically by a pulse injection and detection system \citep{thompson11}, a robust estimator that sums signals while excising one or more extreme stations at each timestep.   The resulting time series is closer to Gaussian. Its noise is estimated from the data and thresholds are adapted periodically to maintain a 7-$\sigma$ cutoff.
However, as our dynamic threshold is not tracked, we cannot use this information to precisely determine $C$ for \vf. We therefore use the conservative estimate of $C=10$ for our limit analysis below.


\subsection{Instrumental broadening and scattering}\label{sec:broadening}
Several pulse broadening effects can reduce the detected signal strength. These interplay between natural broadening effects (i.\,e.\ scattering in the interstellar and intergalactic medium) and instrumental effects. The detected pulse width of an intrinsically unresolved pulse will be the square sum of the intrinsic pulse width and several broadening effects:
\begin{equation}\label{eq:ti}
\tI^2 = 
\ti^2 + 
\tsamp^2 + 
\taus^2 + 
\bigg(\frac{k\,{\rm DM}\,\Delta f}{f^3}\bigg)^2
~,
\end{equation}
where the final term is the ``dispersion broadening'' from a dispersion within finite channel bandwidth $\Delta f$ at frequency $f$, both in MHz, with constant \mbox{$k=8.3\times10^{3}\,{\rm MHz\,cm^{3}\,pc^{-1}}$}. DM refers to the dispersion measure in units of $\dmunits$. The scattering timescale is represented by $\taus$, which typically has a power-law frequency dependence; when combining observations taken at different frequency, as is required for our analysis, one can scale the scattering expected at a common reference frequency such that $\tau_{\rm s} = \tau_{\rm 0}\,(f/f_0)^{\mu}$. For pulsars in the Galaxy and for all scattered FRBs yet observed, $\mu\sim-4$. Note we assume here that detection experiments will search at a variety of pulse widths so that their detection integration time $\simeq t_{\rm eff}$.

The dispersion and scattering of a fast transient can dominate $\tI$ and degrade observational sensitivity. Using the Trott et al.\ framework, we can define a degradation factor, $\epsilon$, which represents the loss from an optimal S/N due to broadening, given intrinsic width $\ti$. Our $\epsilon$ encompasses all losses as reflected in Eq.\,\ref{eq:ti}, not just scattering losses as in \citet{trott13}. The ideal S/N is:
\begin{equation}
{\rm S/N_{opt}} \propto S_{\rm intrinsic}\sqrt{\ti}~.
\end{equation}
When broadening occurs, systems typically find an optimized S/N by integrating longer durations (up to a limit), however this decreases both the noise and the signal at different rates ($\sqrt{t}$ and $t$, respectively) such that the detected S/N after broadening is
\begin{equation}
{\rm S/N}_{\rm broad} \propto \frac{S_{\rm intrinsic}}{t_{\rm eff}/\ti}\sqrt{t_{\rm eff}}~,
\end{equation}
and therefore the degradation factor is simply
\begin{equation}
\epsilon\equiv\frac{\rm S/N_{broad}}{\rm S/N_{opt}} = 
{\rm min}\bigg(1,\sqrt\frac{\ti}{t_{\rm eff}}~\bigg).
\end{equation}
Note that in the above equations we use $\sqrt{w}$ rather than $w$ because the S/N scales as the square root of time. As in \citet{trott13}, our use of S/N, rather than a fluence or flux density, allows us to assess the impact on S/N of different experiments.

\subsection{Two limit scenarios for $\epsilon$}\label{sec:scenarios}
We
present the \vf\ sensitivity in two ways:
\begin{enumerate}
\item {\bf Unknown properties of FRBs.} We apply no assumption about FRB properties or Galactic/extragalactic broadening effects,
so that the results can be compared in the future if FRB properties change with further detections. This equates to using a system with no degradation (i.\,e.~$\epsilon=1$).
\item {\bf Extragalactic FRBs.} Here, we use the average properties of FRBs observed to date, plus a Galactic dispersion and scattering model and varying system set-ups, to determine $\epsilon$ for each \vf\ observation. 
We attempt to correct for the observed latitude-dependence of FRB rates \citep{petroffgb,bannisterbursts} under the assumption that this dependence is due to propagation through the Galactic interstellar medium. 
\end{enumerate}
For the second approach, we use the {\sc ne2001} electron density model \citep{ne2001} to compute scattering and dispersion contributions from the Milky Way, as in \cite{bannisterbursts}. From the FRBs detected to date, we assume the following average values for this analysis:
\begin{equation}
\begin{aligned}
{\rm DM_{\rm exgal}}=~& 816~\dmunits\\
\wfrb=~& 3~{\rm ms}\\
\tau_{\rm exgal}=~& 7~{\rm ms~at~1.0\,GHz}
\end{aligned}
\end{equation}
Here, the ``exgal'' subscript indicates that the quoted quantities exclude the contribution from electrons in the Milky Way itself, which must be added (in a sightline-dependent manner) to obtain the total observed dispersion and scattering. The values above come from ten FRB DM$_{\rm exgal}$ values: 330, 369, 521, 528, 675, 680, 710, 909, 1072, 1553\,$\dmunits$; five FRBs with measured scattering and five with no scattering, projected to 1\,GHz: 3.7, 5.3, 17.7, 17.7, 21.7\,ms; five unscattered FRB widths after accounting for instrumental broadening: 0.57, 3.37, 0.69, 2.87, 6.92\,ms; \citealt{LB,thornton13,bannisterbursts,spitlerFRB,ravishannonFRB,petroffrealtime,thorntonthesis}.
We thus determine $\epsilon$ for each \vf\ observation using the following formula:
\begin{multline}
\epsilon = \sqrt{\wfrb}\Bigg[\tsamp^2 + 
\tau_{\rm exgal}^2(f)+\tau_{\rm MW}^2(f,l,b)
+\bigg(\frac{k\,b\,({\rm DM_{\rm exgal}+DM_{MW}}(l,b))}{f^3}\bigg)^2 \Bigg]^{-1/4}~,
\end{multline}
where again, $\tau_{\rm exgal,MW} = \tau_{\rm 0}\,(f/f_0)^{\mu}$, and we use the {\sc ne2001} Milky Way electron density model to compute scattering and dispersion contributions from the Galaxy. We assume $\mu=-4$ in this analysis. The distribution of $\epsilon$ for each \vf\ observing band is shown in the first panel of Figure\,\ref{fig:data}.


\begin{figure}
\centering
\includegraphics[width=0.5\textwidth,trim=5mm 0mm 0mm 10mm, clip]{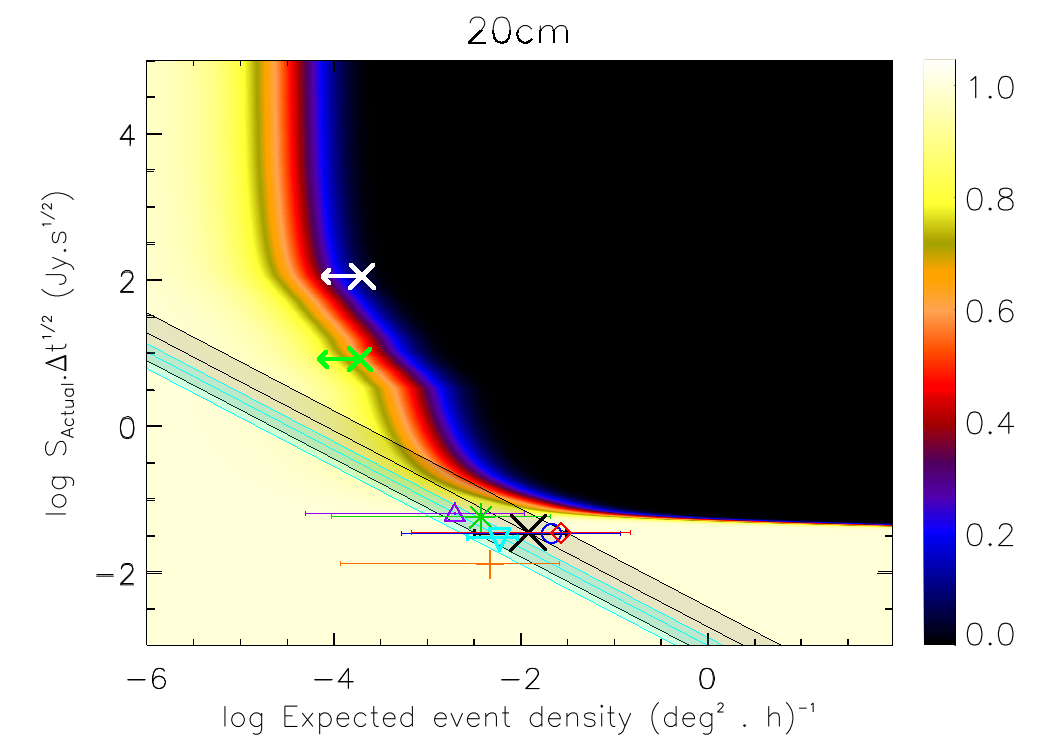}
\caption{\vf\ limits at $\lambda=20$\,cm using the ``extragalactic FRB'' scenario to derive $\epsilon$. The colorbar represents the likelihood that \vf\ would have seen zero events given the event rate and sensitivity metric. The points shown are as in Table\,\ref{table:othersurveys}: \citet{LB} green asterisk, \thorn\ black cross, \citet{spitlerFRB} orange plus, \citet{bannisterbursts} purple triangle, \citet{petroffrealtime} blue circle, \citet{ravishannonFRB} red diamond, \citet{flyseye} upper limit green cross, \citet{championbursts} cyan triangle. A standard $N(>$$S)\propto S^{-3/2}$ scaling is extrapolated from the \thorn\ and Champion et al.\ measurements. The white cross indicates the \vf\ limit that most constrains $\gamma$ (Sec.\,\ref{sec:gamma}).}
\label{fig:20exgal}
\vspace{1mm}
\end{figure}

\begin{figure}
\centering
\includegraphics[width=0.5\textwidth,trim=5mm 0mm 0mm 10mm, clip]{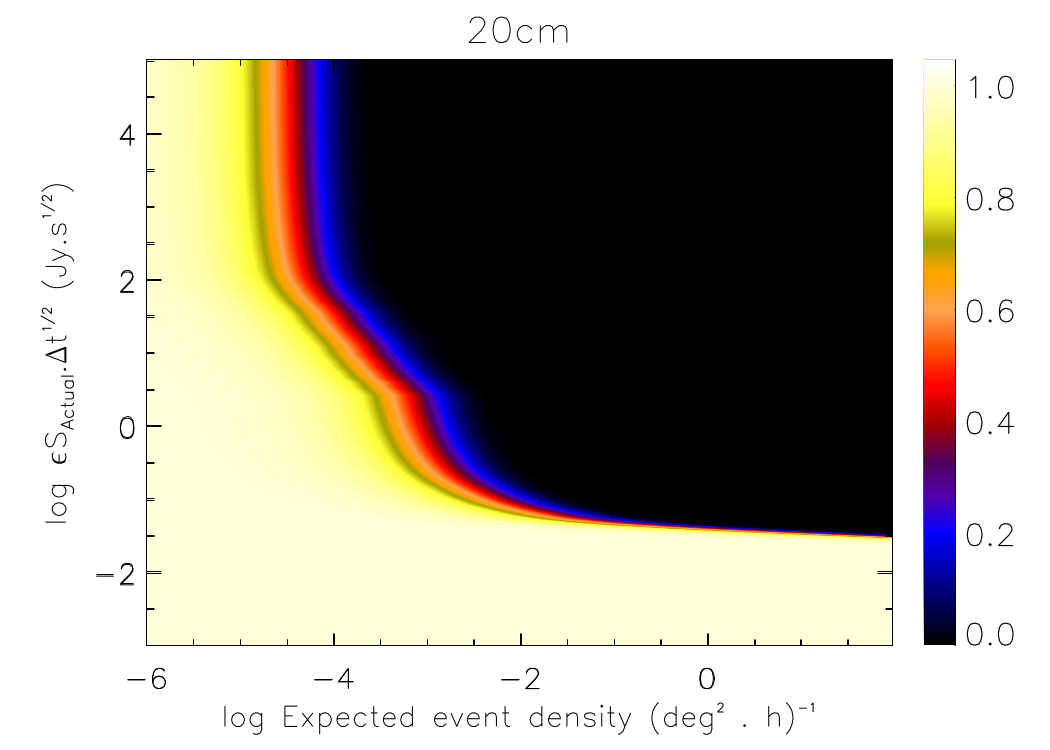}
\caption{\vf\ limits at $\lambda=20$\,cm without any assumptions for $\epsilon$.}\label{fig:20eps1}
\vspace{1mm}
\end{figure}

\begin{figure}
\centering
\includegraphics[width=0.5\textwidth,trim=5mm 0mm 0mm 10mm, clip]{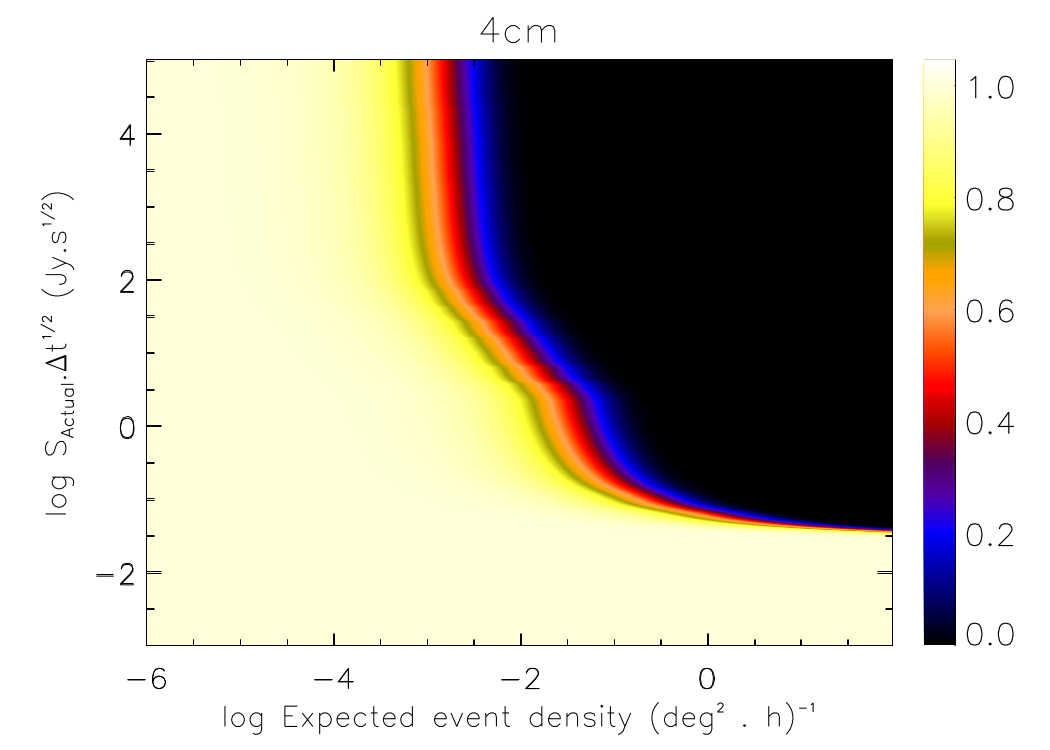}
\caption{\vf\ limits at $\lambda=4$\,cm using the ``extragalactic FRB'' scenario to derive $\epsilon$.}\label{fig:4exgal}
\vspace{1mm}
\end{figure}

\begin{figure}
\centering
\includegraphics[width=0.5\textwidth,trim=5mm 0mm 5mm 10mm, clip]{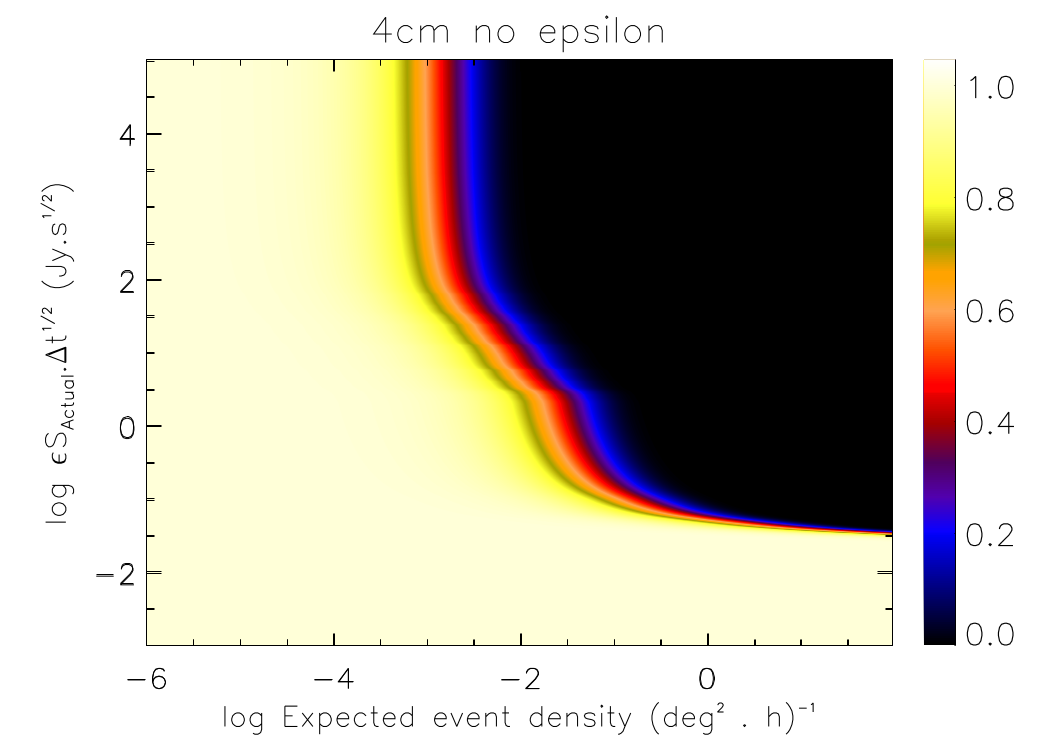}
\caption{\vf\ limits at $\lambda=4$\,cm without any assumptions for $\epsilon$.}\label{fig:4eps1}
\vspace{1mm}
\end{figure}

\section{V-FASTR Limits on FRBs and Millisecond-duration Radio Transients}\label{sec:limitplots}
We combine the inhomogeneous \vf\ data set by forming a 5-dimensional grid of observing band, time sample duration, channel bandwidth, integrated bandwidth, and degradation factor (a collapsed view of this is shown in Fig.\,\ref{fig:data}). Each of these grid cells contains the total on-sky time spent with that combination of observing parameters. The cells of this grid are treated as independent experiments and the net probability (Eq.\,\ref{eq:ptot}) for each limit plot in this section is formed by summing the probabilities over the cells of that band. It is this probability that is plotted in Figures \ref{fig:20exgal}--\ref{fig:4eps1}.

Figures \ref{fig:20eps1} and \ref{fig:4eps1} show the most sensitive bands in \mbox{$\epsilon S_{\rm actual}\sqrt{w}$ - $T_{\rm obs}\times{\rm FOV}$} space using this analysis considering our scenario without assumption of FRB properties. Figures \ref{fig:20exgal} and \ref{fig:4exgal} present the \vf\ limits at the same bands using the ``Extragalactic FRB'' scenario (as described in \S\ref{sec:scenarios}).


\begin{table*}
\centering
  \caption{The sensitivity parameters of surveys that have published FRB discovery rates, and for the ``Fly's Eye'' experiment \citep{flyseye}, which searched for FRBs at 20\,cm. Sensitivity/experimental parameters for most surveys are summarized in \citet{bannisterbursts}. We also use the \citet{spitlerFRB} values assuming their FRB detection was in the sidelobe of the Arecibo antenna, with their reported $S_{\rm min}$ corrected for $\tsky$ and our $\wfrb$. Apart from survey area, the \citet{petroffrealtime} and \citet{ravishannonFRB} parameters are equivalent to those of \thorn. For consistent comparison with the \vf\ limit results, we represent the lower limit of detectable intrinsic FRB flux density for an experiment as $S_{\rm min,opt}=C\cdot{\rm SEFD_{\rm eff}}/\sqrt{{\rm BW}\cdot w_{\rm frb}}$. Calculation uses the gain at the full-width-half-maximum power point of the beam, in concordance with the $\mathcal{T}_{\rm obs}\times {\rm FOV}$ value reported here. Thus in Figure \ref{fig:20exgal} we plot [$S_{\rm min,opt} \sqrt{w_{\rm frb}}/\epsilon$, $N_{\rm FRB}\,/\,(\mathcal{T}_{\rm obs}\times {\rm FOV})$] for each experiment.
  }\label{table:othersurveys}
\begin{tabular}{rccccc}
\hline
{\bf } & $\langle T_{\rm sky} \rangle$ & $S_{\rm min,opt}$
                                          & & $\mathcal{T}_{\rm obs}\times {\rm FOV}$\\
Ref. & (K) & (mJy) &$\langle \epsilon\rangle$ & (h\,deg$^2$) & $N_{\rm FRB}$\\
\hline
\citet{LB}                           & 0.73 & 590 & 0.57 & 272.7 & 1\\ 
\citet{flyseye}                   & 1.65 & $118\times10^3$& 0.76 & 19992 & 0 \\ 
\citet{thornton13} [\thorn]   & 0.85 & 560 & 0.89 & 337.5 & 4\\ 
\citet{spitlerFRB}             & 1.31 & 210 & 0.90 & 215.2 & 1\\ 
\citet{bannisterbursts}& 1.70 & 615 & 0.55 & 509.9 & 1\\ 
\citet{petroffrealtime}      & 0.79 & 555 & 0.89 & 47.5   & 1\\ 
\citet{ravishannonFRB} & 0.64 & 555 & 0.88 & 38.0 & 1\\   
\citet{championbursts} & 0.92& 560& 0.88& 1549 & 10\\
\hline
\vspace{1mm}
\end{tabular}
\end{table*}

\subsection{Comparison with previous FRB detection rates}
Figure \ref{fig:20exgal} compares the rate of several FRB detection experiments with the \vf\ limits derived from 20cm observations. We recalculated the FRB rates for these surveys to provide a consistent analysis with our limits, using values as reported in Table \ref{table:othersurveys}. The most constrained of these measurements come from \citet{championbursts}, who included the ten FRBs detected from the HTRU-S Survey and thus have the smallest Poisson error of the various experiments. This measurement is the most constrained because it reports the largest number of detections from a single survey with uniform parameters.

This figure demonstrates that our limit is currently consistent with all FRB measurements thus far; that is, we had a very low probability of a detection given the intrinsic FRB rates and flux densities implied by these experiments.

%

Nevertheless, our multi-dimensional limits in $S\sqrt{\wfrb}$, $\mathcal{T}_{\rm obs}\times {\rm FOV}$, and frequency space, allow us to statistically constrain the physical parameters of the FRB population, in particular the average spectral index and the index of FRB source counts, as described below.
%

\subsection{FRB Spectral Index Limits}
We now consider physical limits on the observed FRB population to date. For the analysis in this and the next section, we take the 20\,cm detections of \citet[][hereafter \thorn]{thornton13} and \citet[][hereafter C16]{championbursts} as an ``anchor point'' for measured FRB rates. This is a simplifying assumption; further analyses have shown that the error bars and flux scale of FRB rates are highly uncertain and still the subject of ongoing debate \citep{keanepetroff,rane+16}. 
Thus, while a maximum likelihood analysis would be the most rigorous path to derive the implied spectral dependence and number counts for FRBs, particularly if \vf\ were to make a detection, such an analysis is not warranted in a null detection experiment with only two frequencies to fit. Note that T13 searched a subset of C16 data, and T13's reported rate is about double that of C16; we report results from both of these rates to demonstrate how our results change with different FRB rate measurements.

First, we consider limits on the spectrum of FRBs. We use our ``extragalactic FRBs'' scenario to place these limits, as these reflect most accurately any effects of frequency-dependent scattering and instrumental sensitivity. The most simplistic model for a radio spectral density is defined by power-law index $\alpha$, defined with fluence $F\propto f^{\alpha}$. 
We can place a two-point spectral limit on $\alpha$ at each of \vf's single-receiver observing bands. We do this at the 20\,cm event rate surface density of \thorn\ (4/[337.5\,h\,deg$^2$]) and C16 (10/[1549\,h\,deg$^2$]), scaling $\alpha$ until our 5\% probability point in $S\sqrt{\ti}$ meets that of \thorn. We use the 95\% detection contour as a proxy for the 95\% limit on intrinsic source flux density, with the caveat that, in general, these are not the same quantity. This corresponds to the 95\% confidence point in our null hypothesis, $P(H_0)$, that \vf\ should have detected at least one burst given $\alpha$ and the \thorn\ and C16 measurements. These frequency-dependent limits are shown in Table \ref{table:SED}. 

If FRB emission is truly broad-band and has no break in its spectrum, we can state that our limits on spectral index are $-7.6<\alpha<5.8$ if the Champion et al.\ sensitivity is taken at face value. Specifically, our most stringent limits are for the 4\,cm and 90\,cm bands, such that $\alpha_{\rm 20\,cm}^{\rm 4\,cm}<+5.8$ and $\alpha_{\rm 90\,cm}^{\rm 20\,cm}>-7.6$. To our knowledge, the former is the first reported limit on the high-frequency spectral index (upper limit) of the FRB population as a whole. Recently, \citet{artemis} also reported a two-point spectral index lower limit of $\alpha^{\rm 20\,cm}_{\rm 2\,m}>+0.1$.

The $\alpha$ values associated with known radio sources---particularly those with coherent emission as is expected in the case of FRBs---tend to be negative, however for these our limits are not particularly constraining.
For instance, while a few pulsars have flat spectra, they have declining spectra on average $\alpha\simeq-1.4$, and down to $\alpha\simeq-3.5$ \citep[\eg][]{bates13}.

Our limits, however, are interesting in light of two facts. First, one of the few reported FRBs with a spectral index measurement, FRB\,121102, has been found to repeat and has erratic frequency-dependent amplitudes. These induce a huge range of indices measured for different detections, ranging $-$10.4 to +13.6 \citep{repeatingFRBnature,scholz+16}. Our limits rule out that a significant population of bursts commonly have such extreme spectral indices.

Second, \citet{kulkarni14} discuss the effect of free-free absorption of FRB emission by a medium surrounding the FRB progenitor, which would also produce the excess DM observed in FRBs: that is, a non-cosmological FRB scenario where most of the excess DM does not arise from the intergalactic medium. One of the implications of a medium with even a modest optical depth is that the spectrum will be strongly inverted, well above $\alpha\simeq+3.2$ for most bursts. Our upper limit on $\alpha$ thus implies that the average FRB population is not surrounded by compact, photo-ionized nebulae, assuming the region remains optically thick up to at least 6\,GHz.
The implication for this is that the origin of the bulk of FRBs' excessive dispersion measures are on average not source-local if the source is $\mathcal{O}(10,000)$\,K. Some FRB models pinpoint that FRBs arise in nearby galaxies; for these models to hold, therefore, the progenitor sources are implied to be preferentially positioned closer to the centers of the host galaxies, i.\,e.~at a position where the host galaxy's interstellar medium can contribute the large excess DM observed in FRBs.

\begin{table}
\centering
  \caption{The upper limit on power-law spectral index, $\alpha$, for each of \vf's effective center frequencies using the \citet{thornton13} (T13) FRB rate measurement and the \citet{championbursts} (C16) rate measurements. For wavelengths shorter than 4\,cm, the $\mathcal{T}_{\rm obs}\times {\rm FOV}$ product for those bands was too small to allow a comparison with the \thorn\ rate.
}\label{table:SED}
\begin{tabular}{cccc}
\hline
{\bf Band} & {\bf $\langle f_{\rm ctr} \rangle$} &{\bf $\alpha$ (T13)} & {\bf $\alpha$ (C16)}\\
{\bf (cm)} & {\bf (GHz)} &  {\bf limit} & {\bf limit}\\
\hline
90 & 0.318 & $>-6.4$ & $>-7.6$\\
50 & 0.465 & $>-10.8$ &$>-11.5$\\
20 & 1.55   & $<14.1$ & $<18.9$ \\
13 & 2.28   & $<24.1$ & $<33.1$\\
6 &    5.95  & $<4.5$ &  $<5.2$\\
4 &    8.42  & $<4.0$ &  $<5.8$\\
\hline
\end{tabular}
\end{table}

\subsection{FRB Source Count Limits}\label{sec:gamma}
Next, we can derive limits on a power-law scaling of FRB population number counts using our limits in the 20\,cm band. Given our two-point $\alpha$ limits, we make the assumption that the small difference between the average \vf\ and Parkes center frequencies---1.550 and 1.352\,GHz, respectively---is negligible when we place our $\gamma$ limit (i.\,e.\ we assume $\alpha=0$).
As with the spectral index, a typical dependence assumed for this scaling is a power-law, $N(>$$S)\propto S^\gamma$. The standard $\gamma=-3/2$ projection is exhibited in Fig.\,\ref{fig:20exgal} for \thorn\ and C16. Using \vf's 20\,cm limit, we can extrapolate the \thorn\ and C16 measurements to our most sensitive 5\% confidence point, hence placing a limit of $\gamma<-0.5$ and $\gamma<-0.4$, respectively. As in the previous section, we use the 95\% detection contour here as a proxy for the 95\% limit on intrinsic source flux density, with the caveat that in general, these are not the same quantity. We also include in Fig.\,\ref{fig:20exgal} the 20\,cm measurement from the Allen Telescope Array's ``Fly's Eye'' experiment \cite{flyseye}, which provides the most constraining limit to date, of $\gamma<-0.76$.

Both $\gamma$ limits are in concordance with a cosmological population of FRBs in a Euclidean universe, where $\gamma=-1.5$. They also allow sufficient room, for instance, for a non-cosmological population where our sensitivity is source-limited rather than distance-limited \citep[e.\,g.][]{penconnor}. It also still permits the possibility of luminosity or density evolution of FRBs over cosmic distances.

%
%
%

\section{Conclusions}
We have reported constraints on the Fast Radio Burst population based on observations using the \vf\ experiment, which spans observing frequencies from 350\,MHz--90\,GHz. Analysis based on the non-detection of FRBs in \vf\ data to date allows us to draw several conclusions:
\begin{enumerate}
\item Our non-detection at 20\,cm is consistent with the FRB rate based on the detections of \citet{championbursts}. If FRBs are distributed with a $N(>$$S)\propto S^\gamma$ distribution with $\gamma=-1.5$, we find a probability of $\sim$91\% that no FRBs would have been detected by \vf\ data to date (Fig.\,\ref{fig:20exgal}).
\item We have set multi-wavelength constraints on the two-point radio spectral index, $\alpha$. Our most stringent limits are for the 4\,cm and 90\,cm bands, such that $\alpha_{\rm 20\,cm}^{\rm 4\,cm}<$+5.8 and $\alpha_{\rm 90\,cm}^{\rm 20\,cm}>$-7.6. This in particular places constraints on any FRB-local photo-ionized nebula, which would even with a modest optical depth push spectral indices above $\alpha\sim3$ \citep{kulkarni14}.
\item We place a limit on FRB number counts at 20\,cm to be $\gamma<$-0.4. This limit, however, is not more constraining than the limits placed by other experiments on this parameter \citep{flyseye}.
\end{enumerate}

\section{Acknowledgements}
The National Radio Astronomy Observatory is a facility of the National Science Foundation operated under cooperative agreement by Associated Universities, Inc. We acknowledge the exellent commentary by two referees and the ApJ statistician on the paper's manuscripts. This work was carried out in part at the Jet Propulsion Laboratory, California Institute of Technology, under a contract with the National Aeronautics and Space Administration. Parts of this research were conducted by the Australian Research Council Centre of Excellence for All-sky Astrophysics (CAASTRO), through project number CE110001020. CMT is supported by an Australian Research Council DECRA Fellowship through project number DE140100316.

\bibliographystyle{mn2e}
\bibliography{paperbib}

\begin{thebibliography}{}

\bibitem[\protect\citeauthoryear{{Bates}, {Lorimer} \& {Verbiest}}{{Bates}
  et~al.}{2013}]{bates13}
{Bates} S.~D.,  {Lorimer} D.~R.,    {Verbiest} J.~P.~W.,  2013, \mnras, 431,
  1352

\bibitem[\protect\citeauthoryear{{Burke-Spolaor} \&
  {Bannister}}{{Burke-Spolaor} \& {Bannister}}{2014}]{bannisterbursts}
{Burke-Spolaor} S.,  {Bannister} K.~W.,  2014, \apj, 792, 19

\bibitem[\protect\citeauthoryear{{Champion}, {Petroff}, {Kramer}, {Keith},
  {Bailes}, {Barr}, {Bates} \& {et al.}}{{Champion}
  et~al.}{2016}]{championbursts}
{Champion} D.~J.,  {Petroff} E.,  {Kramer} M.,  {Keith} M.~J.,  {Bailes} M.,
  {Barr} E.~D.,  {Bates} S.~D.,    {et al.} 2016, \mnras

\bibitem[\protect\citeauthoryear{{Coenen}, {van Leeuwen}, {Hessels},
  {Stappers}, {Kondratiev}, {Alexov}, {Breton}, {Bilous} \& et al.}{{Coenen}
  et~al.}{2014}]{lofarlimits}
{Coenen} T.,  {van Leeuwen} J.,  {Hessels} J.~W.~T.,  {Stappers} B.~W.,
  {Kondratiev} V.~I.,  {Alexov} A.,  {Breton} R.~P.,  {Bilous} A.,    et al.
  2014, \aap, 570, A60

\bibitem[\protect\citeauthoryear{{Cordes} \& {Lazio}}{{Cordes} \&
  {Lazio}}{2002}]{ne2001}
{Cordes} J.~M.,  {Lazio} T.~J.~W.,  2002, arXiv:0207156

\bibitem[\protect\citeauthoryear{{de Oliveira-Costa}, {Tegmark}, {Gaensler},
  {Jonas}, {Landecker} \& {Reich}}{{de Oliveira-Costa}
  et~al.}{2008}]{globalskymodel}
{de Oliveira-Costa} A.,  {Tegmark} M.,  {Gaensler} B.~M.,  {Jonas} J.,
  {Landecker} T.~L.,    {Reich} P.,  2008, \mnras, 388, 247

\bibitem[\protect\citeauthoryear{{Karastergiou}, {Chennamangalam}, {Armour},
  {Williams}, {Mort}, {Dulwich}, {Salvini}, {Magro} \& et al.}{{Karastergiou}
  et~al.}{2015}]{artemis}
{Karastergiou} A.,  {Chennamangalam} J.,  {Armour} W.,  {Williams} C.,  {Mort}
  B.,  {Dulwich} F.,  {Salvini} S.,  {Magro} A.,    et al. 2015, ArXiv e-prints

\bibitem[\protect\citeauthoryear{{Keane}, {Johnston}, {Bhandari}, {Barr},
  {Bhat}, {Burgay}, {Caleb} \& {et al}}{{Keane} et~al.}{2016}]{keanenatureFRB}
{Keane} E.~F.,  {Johnston} S.,  {Bhandari} S.,  {Barr} E.,  {Bhat} N.~D.~R.,
  {Burgay} M.,  {Caleb} M.,    {et al} 2016, \nat, 530, 453

\bibitem[\protect\citeauthoryear{{Keane} \& {Petroff}}{{Keane} \&
  {Petroff}}{2015}]{keanepetroff}
{Keane} E.~F.,  {Petroff} E.,  2015, \mnras, 447, 2852

\bibitem[\protect\citeauthoryear{{Kulkarni}, {Ofek}, {Neill}, {Zheng} \&
  {Juric}}{{Kulkarni} et~al.}{2014}]{kulkarni14}
{Kulkarni} S.~R.,  {Ofek} E.~O.,  {Neill} J.~D.,  {Zheng} Z.,    {Juric} M.,
  2014, \apj, 797, 70

\bibitem[\protect\citeauthoryear{{Law}, {Bower}, {Burke-Spolaor}, {Butler},
  {Lawrence}, {Lazio}, {Mattmann}, {Rupen}, {Siemion} \& {VanderWiel}}{{Law}
  et~al.}{2014}]{caseyVLA}
{Law} C.~J.,  {Bower} G.~C.,  {Burke-Spolaor} S.,  {Butler} B.,  {Lawrence} E.,
   {Lazio} T.~J.~W.,  {Mattmann} C.~A.,  {Rupen} M.,  {Siemion} A.,
  {VanderWiel} S.,  2014, ArXiv e-prints

\bibitem[\protect\citeauthoryear{{Lorimer}, {Bailes}, {McLaughlin}, {Narkevic}
  \& {Crawford}}{{Lorimer} et~al.}{2007}]{LB}
{Lorimer} D.~R.,  {Bailes} M.,  {McLaughlin} M.~A.,  {Narkevic} D.~J.,
  {Crawford} F.,  2007, Science, 318, 777

\bibitem[\protect\citeauthoryear{{Luan} \& {Goldreich}}{{Luan} \&
  {Goldreich}}{2014}]{luan}
{Luan} J.,  {Goldreich} P.,  2014, \apjl, 785, L26

\bibitem[\protect\citeauthoryear{{Masui}, {Lin}, {Sievers}, {Anderson},
  {Chang}, {Chen}, {Ganguly}, {Jarvis}, {Kuo}, {Li}, {Liao}, {McLaughlin},
  {Pen}, {Peterson}, {Roman}, {Timbie}, {Voytek} \& {Yadav}}{{Masui}
  et~al.}{2015}]{masui}
{Masui} K.,  {Lin} H.-H.,  {Sievers} J.,  {Anderson} C.~J.,  {Chang} T.-C.,
  {Chen} X.,  {Ganguly} A.,  {Jarvis} M.,  {Kuo} C.-Y.,  {Li} Y.-C.,  {Liao}
  Y.-W.,  {McLaughlin} M.,  {Pen} U.-L.,  {Peterson} J.~B.,  {Roman} A.,
  {Timbie} P.~T.,  {Voytek} T.,    {Yadav} J.~K.,  2015, \nat, 528, 523

\bibitem[\protect\citeauthoryear{{McLaughlin}, {Lyne}, {Lorimer}, {Kramer},
  {Faulkner}, {Manchester}, {Cordes}, {Camilo}, {Possenti}, {Stairs} \& et
  al.}{{McLaughlin} et~al.}{2006}]{RRATs}
{McLaughlin} M.~A.,  {Lyne} A.~G.,  {Lorimer} D.~R.,  {Kramer} M.,  {Faulkner}
  A.~J.,  {Manchester} R.~N.,  {Cordes} J.~M.,  {Camilo} F.,  {Possenti} A.,
  {Stairs} I.~H.,    et al. 2006, \nat, 439, 817

\bibitem[\protect\citeauthoryear{{Pen} \& {Connor}}{{Pen} \&
  {Connor}}{2015}]{penconnor}
{Pen} U.-L.,  {Connor} L.,  2015, ArXiv e-prints

\bibitem[\protect\citeauthoryear{{Petroff}, {Bailes}, {Barr}, {Barsdell},
  {Bhat}, {Bian}, {Burke-Spolaor}, {Caleb}, {Champion}, {Chandra}, {Da Costa}
  \& et al.}{{Petroff} et~al.}{2015}]{petroffrealtime}
{Petroff} E.,  {Bailes} M.,  {Barr} E.~D.,  {Barsdell} B.~R.,  {Bhat} N.~D.~R.,
   {Bian} F.,  {Burke-Spolaor} S.,  {Caleb} M.,  {Champion} D.,  {Chandra} P.,
  {Da Costa} G.,    et al. 2015, \mnras, 447, 246

\bibitem[\protect\citeauthoryear{{Petroff}, {van Straten}, {Johnston},
  {Bailes}, {Barr}, {Bates}, {Bhat}, {Burgay}, {Burke-Spolaor}, {Champion} \&
  et al.}{{Petroff} et~al.}{2014}]{petroffgb}
{Petroff} E.,  {van Straten} W.,  {Johnston} S.,  {Bailes} M.,  {Barr} E.~D.,
  {Bates} S.~D.,  {Bhat} N.~D.~R.,  {Burgay} M.,  {Burke-Spolaor} S.,
  {Champion} D.,    et al. 2014, \apjl, 789, L26

\bibitem[\protect\citeauthoryear{{Rane}, {Lorimer}, {Bates}, {McMann},
  {McLaughlin} \& {Rajwade}}{{Rane} et~al.}{2016}]{rane+16}
{Rane} A.,  {Lorimer} D.~R.,  {Bates} S.~D.,  {McMann} N.,  {McLaughlin} M.~A.,
     {Rajwade} K.,  2016, \mnras, 455, 2207

\bibitem[\protect\citeauthoryear{{Ravi}, {Shannon} \& {Jameson}}{{Ravi}
  et~al.}{2015}]{ravishannonFRB}
{Ravi} V.,  {Shannon} R.~M.,    {Jameson} A.,  2015, \apjl, 799, L5

\bibitem[\protect\citeauthoryear{{Scholz}, {Spitler}, {Hessels}, {Chatterjee},
  {Cordes}, {Kaspi}, {Wharton} \& {et al.}}{{Scholz} et~al.}{2016}]{scholz+16}
{Scholz} P.,  {Spitler} L.~G.,  {Hessels} J.~W.~T.,  {Chatterjee} S.,  {Cordes}
  J.~M.,  {Kaspi} V.~M.,  {Wharton} R.~S.,    {et al.} 2016, ArXiv e-prints

\bibitem[\protect\citeauthoryear{{Siemion}, {Bower}, {Foster}, {McMahon},
  {Wagner}, {Werthimer}, {Backer}, {Cordes} \& {van Leeuwen}}{{Siemion}
  et~al.}{2012}]{flyseye}
{Siemion} A.~P.~V.,  {Bower} G.~C.,  {Foster} G.,  {McMahon} P.~L.,  {Wagner}
  M.~I.,  {Werthimer} D.,  {Backer} D.,  {Cordes} J.,    {van Leeuwen} J.,
  2012, \apj, 744, 109

\bibitem[\protect\citeauthoryear{{Spitler}, {Cordes}, {Hessels}, {Lorimer},
  {McLaughlin}, {Chatterjee}, {Crawford}, {Deneva}, {Kaspi}, {Wharton},
  {Allen}, {Bogdanov}, {Brazier}, {Camilo} \& et al.}{{Spitler}
  et~al.}{2014}]{spitlerFRB}
{Spitler} L.~G.,  {Cordes} J.~M.,  {Hessels} J.~W.~T.,  {Lorimer} D.~R.,
  {McLaughlin} M.~A.,  {Chatterjee} S.,  {Crawford} F.,  {Deneva} J.~S.,
  {Kaspi} V.~M.,  {Wharton} R.~S.,  {Allen} B.,  {Bogdanov} S.,  {Brazier} A.,
  {Camilo} F.,    et al. 2014, \apj, 790, 101

\bibitem[\protect\citeauthoryear{{Spitler}, {Scholz}, {Hessels}, {Bogdanov},
  {Brazier}, {Camilo}, {Chatterjee} \& {et al.}}{{Spitler}
  et~al.}{2016}]{repeatingFRBnature}
{Spitler} L.~G.,  {Scholz} P.,  {Hessels} J.~W.~T.,  {Bogdanov} S.,  {Brazier}
  A.,  {Camilo} F.,  {Chatterjee} S.,    {et al.} 2016, ArXiv e-prints

\bibitem[\protect\citeauthoryear{{Thompson}, {Wagstaff}, {Brisken}, {Deller},
  {Majid}, {Tingay} \& {Wayth}}{{Thompson} et~al.}{2011}]{thompson11}
{Thompson} D.~R.,  {Wagstaff} K.~L.,  {Brisken} W.~F.,  {Deller} A.~T.,
  {Majid} W.~A.,  {Tingay} S.~J.,    {Wayth} R.~B.,  2011, \apj, 735, 98

\bibitem[\protect\citeauthoryear{{Thornton}, {Stappers}, {Bailes}, {Barsdell},
  {Bates}, {Bhat}, {Burgay} \& et al.}{{Thornton} et~al.}{2013}]{thornton13}
{Thornton} D.,  {Stappers} B.,  {Bailes} M.,  {Barsdell} B.,  {Bates} S.,
  {Bhat} N.~D.~R.,  {Burgay} M.,    et al. 2013, Science, 341, 53

\bibitem[\protect\citeauthoryear{{Thornton}}{{Thornton}}{2013}]{thorntonthesis}
{Thornton} T.,  2013, PhD thesis, University of Manchester (UK)

\bibitem[\protect\citeauthoryear{{Trott}, {Tingay}, {Wayth}, {Thompson},
  {Deller}, {Brisken}, {Wagstaff}, {Majid}, {Burke-Spolaor}, {Macquart} \&
  {Palaniswamy}}{{Trott} et~al.}{2013}]{trott13}
{Trott} C.~M.,  {Tingay} S.~J.,  {Wayth} R.~B.,  {Thompson} D.~R.,  {Deller}
  A.~T.,  {Brisken} W.~F.,  {Wagstaff} K.~L.,  {Majid} W.~A.,  {Burke-Spolaor}
  S.,  {Macquart} J.-P.~R.,    {Palaniswamy} D.,  2013, \apj, 767, 4

\bibitem[\protect\citeauthoryear{{Vedantham}, {Ravi}, {Mooley}, {Frail},
  {Hallinan} \& {Kulkarni}}{{Vedantham} et~al.}{2016}]{vedantham16}
{Vedantham} H.~K.,  {Ravi} V.,  {Mooley} K.,  {Frail} D.,  {Hallinan} G.,
  {Kulkarni} S.~R.,  2016, ArXiv e-prints

\bibitem[\protect\citeauthoryear{{Wayth}, {Brisken}, {Deller}, {Majid},
  {Thompson}, {Tingay} \& {Wagstaff}}{{Wayth} et~al.}{2011}]{wayth11}
{Wayth} R.~B.,  {Brisken} W.~F.,  {Deller} A.~T.,  {Majid} W.~A.,  {Thompson}
  D.~R.,  {Tingay} S.~J.,    {Wagstaff} K.~L.,  2011, \apj, 735, 97

\bibitem[\protect\citeauthoryear{{Wayth}, {Tingay}, {Deller}, {Brisken},
  {Thompson}, {Wagstaff} \& {Majid}}{{Wayth} et~al.}{2012}]{wayth12}
{Wayth} R.~B.,  {Tingay} S.~J.,  {Deller} A.~T.,  {Brisken} W.~F.,  {Thompson}
  D.~R.,  {Wagstaff} K.~L.,    {Majid} W.~A.,  2012, \apjl, 753, L36

\bibitem[\protect\citeauthoryear{{Williams} \& {Berger}}{{Williams} \&
  {Berger}}{2016}]{wbarxiv}
{Williams} P.~K.~G.,  {Berger} E.,  2016, ArXiv e-prints

\bibitem[\protect\citeauthoryear{{Williams}, {Berger} \& {Chornock}}{{Williams}
  et~al.}{2016}]{wbatel}
{Williams} P.~K.~G.,  {Berger} E.,    {Chornock} R.,  2016, The Astronomer's
  Telegram, 8752

\end{thebibliography}

\end{document}